# Simulating spiraling bubble movement in the EL approach

Andreas Weber[a], Hans-Jörg Bart[a], Axel Klar[b]

[a]Chair of Separation Science and Technology, TU Kaiserslautern, PO box 3049
andreas.weber@mv.uni-kl.de; bart@mv.uni-kl.de

[b]Chair of Technomathematics, TU Kaiserslautern, PO box 3049
klar@mathematik.uni-kl.de

**Abstract**

Simulating the detailed movement of a rising bubble can be challenging, especially when it comes to bubble path instabilities. A solution based on the Euler Lagrange (EL) approach is presented, where the bubbles show oscillating shape and/or instable paths while computational cost are at a far lower level than in DNS. The model calculates direction, shape and rotation of the bubbles. A lateral force based on rotation and direction is modeled to finally create typical instable path lines. This is embedded in an EL simulation, which can resolve bubble size distribution, mass transfer and chemical reactions. A parameter study was used to choose appropriate model constants for a mean bubble size of 3 mm. To ensure realistic solution, validation against experimental data of single rising bubbles and bubble swarms are presented. References with 2D and also 3D analysis are taken into account to compare simulative data in terms of typical geometrical parameters and average field values.

**Keywords**

*Euler Lagrange; bubble column; bubble path instability; ellipsoidal bubble; bubble shape; Zigzag path*

**Introduction**

Understanding bubbles path instabilities is a major challenge since the 1960s. Early measurements were performed by Aybers & Tapucu (1969), where a single bubble was rising in a fluid showing different path lines, like zigzag and spiraling. Nowadays, a 3D camera setup enables a more detailed view (Shew & Pinton, 2006b) and sophisticated simulations (Mougin & Magnaudet, 2006) could help to explain the complex rising behavior. The origin of this behavior is in the turbulent eddies induced behind the bubble during its rise. Instabilities in these eddies lead to an eccentric force on the bubble, which then leads to a deviation from a straight rise resulting into a spiraling or zigzagging path. The development of the turbulent eddy behind a bubble is connected also to its shape and size. Different types of bubble paths can be experienced, based on the bubble size or, in terms of turbulence intensity, on the bubble Reynolds number.

Simulating bubble hydrodynamics is an issue since many years. The most common approach to do this is the simplified Euler Euler (EE) method, where single bubbles are no longer resolved but a bubble number density is taken into account instead. This leads to a much simpler way to calculate for an instable bubble path, where a diffusion term is used, which gives a high loss of detail while the overall spatial distribution of bubbles can be forecast with adequate precision. The other extreme is a high detail direct numerical simulation (DNS) of a single bubble or rather a small bubble swarm. Bubbles are resolved in full detail including the hydrodynamics inside. In doing so, path instabilities can be simulated consequently on the lowest level of scale but at a high computational load.

Free rising ellipsoidal bubbles not only move in straight lines but can describe sinusoidal, zigzag or spiraling paths. The common Euler Euler (EE) simulation techniques can no longer resolve the actual movement patterns and Direct Numerical Simulations (DNS) tend to be very costly when simulating a larger number of bubbles. This work presents a solution to calculate the orientation and shape of bubbles using the Euler Lagrange (EL) approach. Advantages lie within the fast computation and the high level of detail. In comparison to DNS the insides of the bubbles are not calculated in full detail but macroscopic models are employed. Every bubble is calculated individually, having its own size, direction and shape. The surrounding fluid will influence not only the bubble's movement but also the rotation and shape. The actual calculation of the turbulent eddies behind the bubble will not be carried out, but an oscillation orientation model is used and model parameters are calibrated from experimental data.

**EL modeling**

The general EL model describes bubbles as a point volume acting under Newtonian dynamics. Forces created by the surrounding fluid and neighboring bubbles accelerate these Lagrangian points through the domain. The continuous phase itself is calculated using Navier-Stokes equations and can be coupled to the interaction forces by a source term. All forces are calculated for each bubble individually, which produces an individual path for each bubble. Therein lies one of the advantage of the EL approach in comparison to the EE methods. Bubbles can coalesce and break, which gives a bubble size distribution, with even more detail than common method of moments approaches. Downside of the EL approach is the higher computational load, which is strongly dependent on the number of bubbles simulated. Nevertheless, the EL approach has been used frequently in several different simulations of bubbly flow (Gruber et al., 2015).

**Liquid phase hydrodynamics**

The continuous phase is assumed to be incompressible, basis for calculation is a modified Navier-Stokes equation.

$$\rho \left(\frac{\partial \vec{u_c}}{\partial t} + (\vec{u_c} \cdot \nabla)\vec{u_c}\right) = -\nabla p + \mu \Delta \vec{u_c} + \vec{f} \tag{1}$$

Given is the continuous velocity $u_c$, the pressure $p$, the density $\rho$, the viscosity $\mu$ and the source term $f$. The source term $f$ depicts the forces of the bubbles and will thereby serve as a coupling of the phases. Turbulence is computed using the standard RANS k-epsilon model (Launder & Spalding, 1972) with additional bubble induce turbulence (BIT) (Rzehak & Krepper, 2013). Each bubbles' drag force induces turbulent energy and dissipation in the associated computational grid cell. For each cell this sums up and is added to the source term in the turbulence model.

$$S_k = |\sum \vec{F_D}| \, |\vec{u_c} - \vec{u_b}| \tag{1}$$

$$S_\varepsilon = \frac{C_\varepsilon S_k}{\tau} \tag{2}$$

$$\tau = \frac{d}{\sqrt{k}} \; ; \; C_\varepsilon = 1.0 \tag{3}$$

Here, $S_k$ denotes the source term for the turbulent energy $k$, $F_D$ stands for the drag force. $S_\varepsilon$ is the turbulent dissipation with $\tau$, the turbulent time scale.

**Bubble hydrodynamics**

Bubbles are modeled as point volumes acting under Newtonian dynamics. Their movement is calculated using a number of different forces.

$$m_b \, d\vec{u_b} = dt \sum \vec{F} \tag{4}$$

The sum of forces $\Sigma F$ consists of the buoyancy and weight force $F_B$, the drag force $F_D$, the lift force $F_L$, the virtual mass force $F_{VM}$, the wall lubrication force $F_W$ and the bubble dispersion force $F_{BD}$. Here the subscripts $b$ and $c$ stand for the bubble and the continuous phase accordingly, the subscript $rel$ identifies the relative differences between them. Furthermore, $g$ denotes the gravitational acceleration, $\rho$ stands for densities, $u$ for velocity, $V$ for the bubble's volume, $k$ for the turbulent kinetic energy and $\alpha$ for the phase fraction. Appropriate model parameters are denoted with $C_i$.

$$\vec{F_B} = m_b \vec{g} \left(1 - \frac{\rho_c}{\rho_b}\right) \tag{5}$$

$$\vec{F_L} = m_b \frac{\rho_c}{\rho_b} C_L \vec{u_{rel}} \times \nabla \times \vec{u_c} \tag{6}$$

$$\vec{F_D} = \frac{3}{4} C_D \frac{m_b \rho_c}{d_b \rho_b} |\vec{u_{rel}}| \vec{u_{rel}} \tag{7}$$

$$\vec{F_{VM}} = -C_{VM} \rho_c V_b \left(\frac{D_b \vec{u_b}}{Dt} - \frac{D_c \vec{u_c}}{Dt}\right) \tag{8}$$

$$\vec{F_W} = -C_W V_b \alpha_b \rho_c \vec{u_{rel}}^2 n_{wall} \tag{9}$$

$$\vec{F_{BD}} = -C_{BD} \rho_c k_c \nabla \alpha_b \tag{10}$$

$D_i/Dt$ in eq. (8) denotes the material derivative, meaning that the derivative is made while following the bubble. Drag and lift force coefficients $C_D, C_L$ are calculated using the models of Tomiyama (2004). The virtual mass coefficient is set to $C_{VM} = 0.5$ according to Delnoij et al. (1997), the coefficient for the dispersion force is set to $C_{TD} = 0.1$ (Lahey et al., 1993). This dispersion describes the ambition of bubbles to spread due to collisions with other bubbles. Additionally, a second turbulent dispersion is used to model the collision of bubbles and turbulent eddies. The Random Dispersion Model (Smith & Milelli, 1998) is used to calculate eddies according to the surrounding level of turbulence. Assuming an isotropic turbulence, eddies are traveling through the liquid in a uniformly random direction with a specific life time. In the model, the turbulent eddy lifetime $t_E$ is evaluated, after which a new eddy is calculated.

$$t_E = \left(\frac{3}{2}\right)^{0.5} C_\mu^{\frac{3}{4}} \frac{k}{\varepsilon}; C_\mu = 0.09 \qquad (11)$$

Then the movement direction of the eddy is uniformly chosen while its velocity follows a normal distribution with a variance dependent on the turbulent energy $k$.

$$|\overline{\overline{u_T}}| = N\left(\sqrt{\frac{2}{3}k}; 0\right) \qquad (12)$$

This turbulent velocity is added to the underlying continuous phase velocity for each bubble individually, which is then used to calculate the different bubble forces. Especially the drag force calculation is a crucial step to produce dispersion of the bubbles. In order to achieve the same amount of bubble dispersion like in the experiment, the turbulent velocity for drag force calculation had to be slightly increased.

Change in bubble volume due to the pressure drop while rising is calculated assuming an ideal gas. Mass transfer and/or chemical reactions are not considered in this simulation.

**Ellipsoidal bubble model**

According to the state of the art of bubble simulations almost any model assumptions are based on bubbles having a spherical shape. Simplest example is the Sauter diameter $d_{32}$, which maps the mean volume/surface area ratio of a bubble population to one spherical bubble size. Collision frequency or rather probability, e.g. by O'Rourke (1981), is based on the calculation of the overlapping volume of spheres. Other examples are mass transfer calculation (spherical transport area), coalescence and break-up (interfacial energy of spheres) or simply the distance calculation between bubbles or wall and bubble. Only exception is the drag force, which is often modeled with respect to the shape of the bubble e.g. by Tomiyama (2004) but uses no information about orientation or rotation of the bubble. One has to admit, that the assumption of bubbles being spheres is adequate for many problems. For example, the deviation of the collision probability is negligible if a spatially uniform random orientation of the bubbles is assumed. Nevertheless, for the computation of realistic bubble movement, a better model for the bubble shape has to be chosen.

In this work, the deformed bubble will be approximated with an oblate spheroid, an ellipsoid with two different axes $a$ and $c$, where $c > a$. For simplicities sake, all following figures will contain a simplified 2D version of the spheroid in Figure 1.

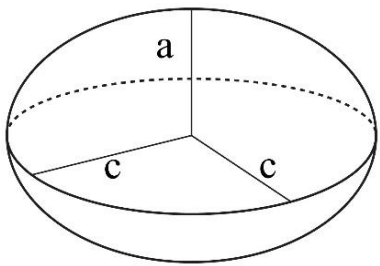

Figure 1: oblate spheroid

The ratio of the axes is chosen to describe the bubble shape via the shape factor *sf*.

$$sf = \frac{1}{\chi} = \frac{a}{c} \tag{13}$$

A shape factor of $sf = 1$ describes a perfect spherical shape, while a lower value stands for a more deformed sphere. $sf = 0$ would describe an infinitely thinned spheroid. Some models are based on empirical measurements using dimensionless numbers for the calculation, others are physical models derived from the interfacial tension and pressure distribution. It turned out that a good result for the simulation problem was achieved by Moore (1965) with the underlying equation:

$$We(\chi) = 4\,\chi^{-\frac{3}{4}}(\chi^3 + \chi - 2)\left[\chi^2 \operatorname{asec}(\chi) - (\chi^2 - 1)^{\frac{1}{2}}\right]^2 (\chi^2 - 1)^{-3} \tag{14}$$

Since the shape factor has to be computed as a function of the Weber number, an approximation has been used:

$$sf = \frac{1}{\operatorname{atanh}\left(\frac{We}{We_{crit}}\right) + 1} \tag{15}$$

Note that this equation is based on the dimensionless Weber number which can represent changes in the shape induced by fluctuating relative velocities. The critical Weber Number $We_{crit} = 3.745$ describes the transition to irregular bubble shapes. If the current Weber number is higher than $We_{crit}$ a shape factor of $sf = 0.2$ is chosen.

$$We = \frac{d\,\rho\,u_{rel}^2}{\sigma} \tag{16}$$

Implementing the new bubble shape into the CFD framework necessitates further usage of the diameter definition of a spherical bubble. The volume of the sphere should be equal to the oblate spheroid volume, which leads to the following basic equations:

$$V_{sphere} = V_{ellipsoid} \tag{17}$$

$$\frac{1}{6}\pi d^3 = \frac{1}{6}\pi a c^2 \tag{18}$$

$$c = \sqrt[3]{\frac{d^3}{sf}} \; ; a = \sqrt[3]{d^3 sf^2} \tag{19}$$

Differently than with common rotation calculation, using the moment of inertia and torque action (Shew & Pinton, 2006a), an approach originally developed by Taylor (1923) is used. The direction of the bubble is described by a vector $p$ pointing in the direction of axis $a$. The change in direction is given as the vector $\dot{p}$, which is deduced from the rotation vector $\omega$ (s. Figure 2). This notation is beneficial because there is no need for a change to spherical coordinates ($\Theta$, $\Phi$), which would lead to higher computational load. The only requirement is a sufficiently small change of the vector $p$, which is fulfilled within the computed time step of the simulation.

$$\dot{\vec{p}} = \vec{\omega} \times \vec{p} \tag{20}$$

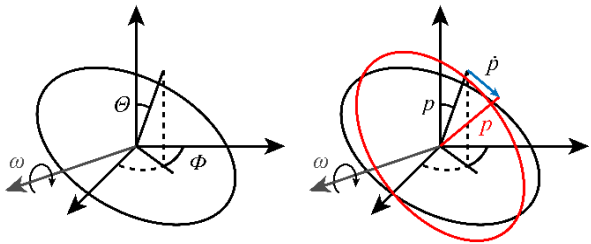

Figure 2: left: orientation in spherical coordinates; right: orientation with p and change in direction $\dot{p}$

This rotation notation has been used by Junk & Illner (2007) to calculate the orientation of rigid ellipsoidal bodies in a Stokes flow. Based on these equations, a modified model for ellipsoidal bubbles was developed. In general, the bubble rotation is calculated using an explicit Euler algorithm:

$$d\vec{p} = \dot{\vec{p}} dt \tag{21}$$

$$d\dot{\vec{p}} = \bar{\gamma}(\vec{G} - \dot{\vec{p}})dt + R \overrightarrow{p_T} dt \tag{22}$$

Just as the bubble's position and velocity, the orientation $p$ changes due to the orientation change $\dot{p}$. The bubble's rotation relaxes against the outer rotation $G$ with the factor $\bar{\gamma}$. Here, $\bar{\gamma}$ denotes a simplified interaction of the bubble's moment of inertia and torque. A low $\bar{\gamma}$ implies a high bubble mass/inertia and thus a slower rotation due to outer forces. The additional term $\overrightarrow{p_T}$ resembles a random rotation due to turbulence, where $R$ is used to scale the effect of turbulent randomness. It is generated similarly to the turbulent dispersion velocity $\overline{u_T}$. Since the effective moment of inertia of a gas bubble inside a fluid is unknown, $\bar{\gamma}$ has to be derived from experimental data.

$$\vec{G} = J_1 \frac{1}{2}(rot\,\overrightarrow{u_C}) \times \vec{p}$$

$$+ J_2\,\lambda(\,S[\overrightarrow{u_C}]\vec{p} - (\vec{p}^T S[\overrightarrow{u_C}]\,\vec{p})\,\vec{p}) \tag{23}$$

$$+ J_3(-\vec{g} - \vec{p})$$

$$\lambda = \frac{sf^2 - 1}{sf^2 + 1} \tag{24}$$

The outer rotation $G$ is a summation of three main mechanics acting on the body (eq. 23). The first line implies that the rotation of the surrounding fluid is transmitted to the bubble itself. The second line describes the rotation induced by shear stress in the surrounding fluid, $S[\overrightarrow{u_C}]$ is the symmetric part of the Jacobian matrix.

$$S[\overrightarrow{u_C}] = \frac{1}{2}(\nabla \cdot \overrightarrow{u_C} + (\nabla \cdot \overrightarrow{u_C})^T) \tag{25}$$

The third line is an addition to the original model (Junk & Illner, 2007) and describes the ambition of the gas bubble to orient itself along the gravitation direction $g$. Since the original equation was deduced for rigid bodies only, all of the above-mentioned mechanics are weighted by the parameters $J_i$ to fit experimental data of (non-rigid) bubbles. This also compensates for the fact, that there is no Stokes flow around the bubble and its interface is mobile (slip condition).

Especially the third line of eq. 23 leads to an oscillatory system, which enables the bubble to describe sinusoidal or helical orientation characteristics. While the orientation of the bubble is changing, the vector of orientation change $\dot{p}$ will eventually point to a direction not perpendicular to $p$, thus rising in size and finally damping and stopping the oscillation. To prevent this, the direction change vector will be moved to the plane normal to the orientation vector by subtracting the part parallel to it:

$$\dot{\vec{p}}' = \dot{\vec{p}} - (\vec{p}(\vec{p} \cdot \dot{\vec{p}})) \tag{26}$$

In total, this will lead to a slight damping of the oscillation which can be eliminated by preserving the magnitude of the orientation change vector:

$$\dot{\vec{p}}'' = \dot{\vec{p}}' \frac{|\dot{\vec{p}}|}{|\dot{\vec{p}}'|} \tag{27}$$

To preserve robust behavior a slight damping ($\delta = 0.2$) is executed for the orientation change:

$$\dot{\vec{p}}''' = \dot{\vec{p}}'' \, e^{-dt\,\delta} \tag{28}$$

Finally, with an interaction force based on the orientation, the bubble experiences a drift perpendicular to the main movement direction. This results in a bubble trajectory describing helical and sinusoidal (zigzag) paths. This perpendicular force has the same direction as the change in direction (Mougin & Magnaudet, 2006), which gives:

$$\dot{\vec{p}} \sim \vec{F}_S \tag{29}$$

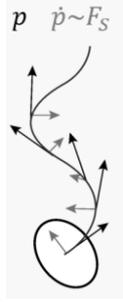

Figure 3: Direction of forces and direction change on a sinusoidal bubble path

This force is modeled as a modified drag force. Also, to imitate real bubble behavior, we limit the force to only occur on bubbles within a certain range of Reynolds number (Aybers & Tapucu, 1969). This will stop the oscillating motion for very small ($d < 0.8$ mm) and large bubbles ($d > 3.5$ mm) as observed in experiments.

$$\vec{F}_S = \begin{cases} \beta |\vec{F}_D| \frac{\dot{\vec{p}}}{|\dot{\vec{p}}|} \sqrt{|\dot{\vec{p}}|} & if \ Re = [500, 1300] \\ 0 & else \end{cases} \tag{30}$$

The additional side force due to bubble rotation is modeled with the magnitude of the current drag force $F_D$ and the bubble rotation but will point in the direction of $\dot{p}$ instead. (It turned out, that a scaling linear to the magnitude of the direction change $|\dot{p}|$ will lead to instabilities easily, which can be stabilized by using the square root $\sqrt{|\dot{p}|}$ instead.) The bubble path amplitude is calibrated with the parameter $\beta$, where a higher value implies a larger amplitude of the resulting oscillating path.

**Parameter estimation**

Parameters $\bar{\gamma}$, R and $J_i$ (eqs. (22), (23)) are derived from a parameter study and with analysis of the oscillatory equations. The characteristic bubble path amplitudes, frequency and wavelength shown in Shew & Pinton (2006b) were taken to set first limits to the parameter study. In a more detailed analysis, the parameters are correlated to experimental data showing bubble swarms rather than single rising bubbles. The resulting simulated bubble paths are then compared to experimental data from the Bubble Column Reactor Database of the University of Magdeburg[1]. Characteristic oscillatory dimensions can be derived from eq. (22) and the third line of eq. (23). This will lead to a simple harmonic oscillator equation with the following form:

$$\ddot{\vec{p}} + \omega_0^2 \vec{p} = 0 \tag{31}$$

with

---

[1] http://www.uni-magdeburg.de/isut/LSS/spp1740/

$$\ddot{\vec{p}} = \bar{\gamma} J_3 \vec{p} \tag{32}$$

Without damping of the oscillation, the frequency $f_0$ can be calculated with the parameters $J_3$ and $\bar{\gamma}$.

$$f_0 = \frac{\omega_0}{2\pi} = \frac{\sqrt{\bar{\gamma} J_3}}{2\pi} \tag{33}$$

It turned out, that this is the case for our oscillatory system, but only if the random rotation is set to zero. A case without random rotation was set up for this reason, the results are shown in Figure 4. Over a wide range, the simulated frequency is identical to the analytical solution for the undamped oscillation.

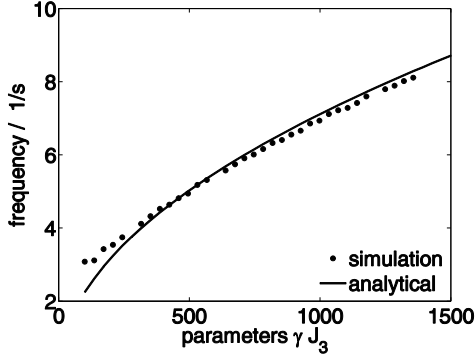

Figure 4: Oscillation parameters vs. bubble path frequency

Most references are made on a basis of a stagnant liquid phase, such that the bubble rise velocity $u_b$ can be easily estimated. This makes it easy to calculate an appropriate wavelength of the generated/observed bubble path using its characteristic frequency.

$$\lambda = \frac{u_b}{f} \tag{34}$$

In case of a non-stagnant liquid, like in a bubble column, the bubble velocity has to be identified at first. It is therefore important to mention, that the wavelength of a bubble path in a dynamic system will differ from most reference experiments, which are made using a well-defined surrounding. Also, in some areas of the column a downward flow occurs, lowering the bubble's rising velocity. To overcome this problem, the bubble velocity and/or wavelengths are averaged over a large number of bubble paths.

**Experimental setup**

For the parameter study, simulation results were compared to experimental measurements from the *Institut für Strömungstechnik und Thermodynamik*, *University of Magdeburg* (Hlawitschka et al., 2016; Zähringer et al., 2014). The experimental setup consists of a cylindrical air-water bubble column with an inner diameter of 14.22 cm and a filling height of 73 cm. The air inlets are positioned at the bottom of the column and consist of four nozzles in a row. The distance between nozzles is 2.2 cm. The measurements were made with a camera system facing the four nozzles side by side (front view) and were also done a second time when facing the nozzles behind each other (side view). With this setup, the bubble positions and velocities were captured so that the trajectories could be reconstructed. With an air throughput of 10 l/h a mean bubble size of $d = 3$ mm was estimated using the same camera setup.

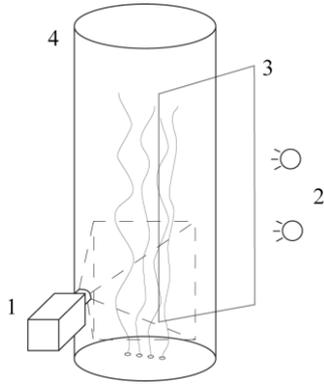

Figure 5: Experimental setup; 1: digital camera(s), 2: light source (laser or LED), 3: light sheet optic, 4: bubble column; from: http://www.uni-magdeburg.de/isut/LSS/spp1740/

An automated analysis was used to estimate the characteristic path shapes, given by the mean and standard deviation of the wavelength $\lambda$ and amplitude $A$ of all bubble paths. To ensure a correct analysis, only those trajectories were considered, that were of appropriate length (minimum of 30 measured positions). Unfortunately, a 3D reconstruction was not possible for the bubble paths, because the side view measurements were not made simultaneously with the front view.

**Computational setup**

The corresponding computational mesh was created using a rectangular grid of 28 x 28 x 146 cells with overlapping cells being removed and reshaped to fit the cylindrical shape. This results in a mesh with 90000 cells, shown in Figure 6. Mesh resolution was intentionally left at approximately 0.125 cm³ per cell (mean cell length of 0.5 cm), because the bubble size must be smaller than the cell size in our EL approach. Cells near the walls are not rectangular anymore, but are also solemnly populated by bubbles due to wall lubrication force. The main bubble flow therefore happens in the center part, where the cells are in perfect rectangular shape.

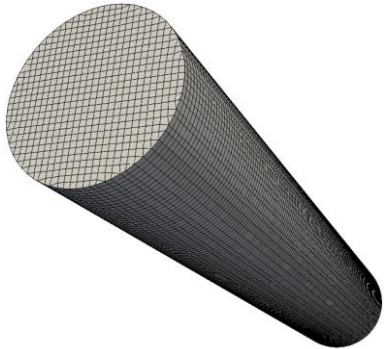

Figure 6: Computational mesh

For boundary conditions (s. Table 1), the simplest possible case could be adopted. All walls except the top outlet patch are combined and have the same properties. Since bubbles are injected using the Lagrangian approach, there is no need for an inlet patch. Instead of a free surface at the top, a slip condition has been set. This approach enables to use a single phase solver but is surely not suitable for estimating hydrodynamics near the surface. Since our interest lies within the lower part of the column, this error is acceptable.

Table 1: *Boundary conditions*

| patch | liquid velocity | pressure | k / epsilon |
|---|---|---|---|
| walls | no-slip condition | zero gradient | zero gradient |

| outlet | slip condition | ambient pressure | zero gradient |
|---|---|---|---|
| internal field, start value | (0 0 0) | ambient pressure | 0.001 / 0.0001 |

On the basis of the experimental measurements, the mean inlet diameter of the bubbles was set to $d = 3$ mm. With a throughput of 10 l/h this leads to a total number of 196.4 bubbles per second to inject from the four nozzles. Bubble breakup, coalescence and mass transfer was turned off in this case, since only the bubble movement is of concern and measurements showed only a minimum of bubble size variations during rise. The experimental measurement area is positioned in the lower 312 mm of the column only, but the whole height of the column was simulated to achieve the correct flow pattern.

Since there are 4 parameters in total to be calibrated, an automated parameter study software has been used (Dakota[2]) to simulate a variety of combinations. The target function was the characteristic bubble flow path parameters, which were compared by screenshots at a first sight. After choosing the most promising parameter interval, a detailed analysis of bubble path frequency and amplitude followed. Therefore, an automated analysis of the generated path lines was used to calculate for mean wavelength and amplitude (~600 trajectories per simulation). The same method has been used to characterize the experimental measurements. Since the experimental data was collected using a 2D visual acquisition, simulation results were also calculated using a 2D mapping. Additionally, a 3D analysis was made and compared to other reference measurements.

---

[2] *https://dakota.sandia.gov/*

## Results and discussion

*Velocity profiles*

Simulated liquid velocity and its deviation are compared to the experimental measurements in Figure 7. The upper figure shows the mean vertical velocity of experiment (left) and simulation (right) on the middle plane of the column. The white line represents the zero value, making it easier to spot the upward and downward flow area. The scale depicted on the left side shows the detailed line plot positions (A-D) in the lower figures. The liquid velocity profile from the experimental measurements shows a slower flow than the simulation produced. Probable reason for this could lie within the modifications done to the drag force calculations in order to acquire the special bubble movement. It has not yet been optimized in order to sustain an appropriate liquid velocity. Also, the liquid phase resolution is quite low in order to maintain an appropriate aspect ratio of bubble length to cell length. Aim of this work lies within the simulation of correct bubble paths rather than optimal liquid phase hydrodynamics.

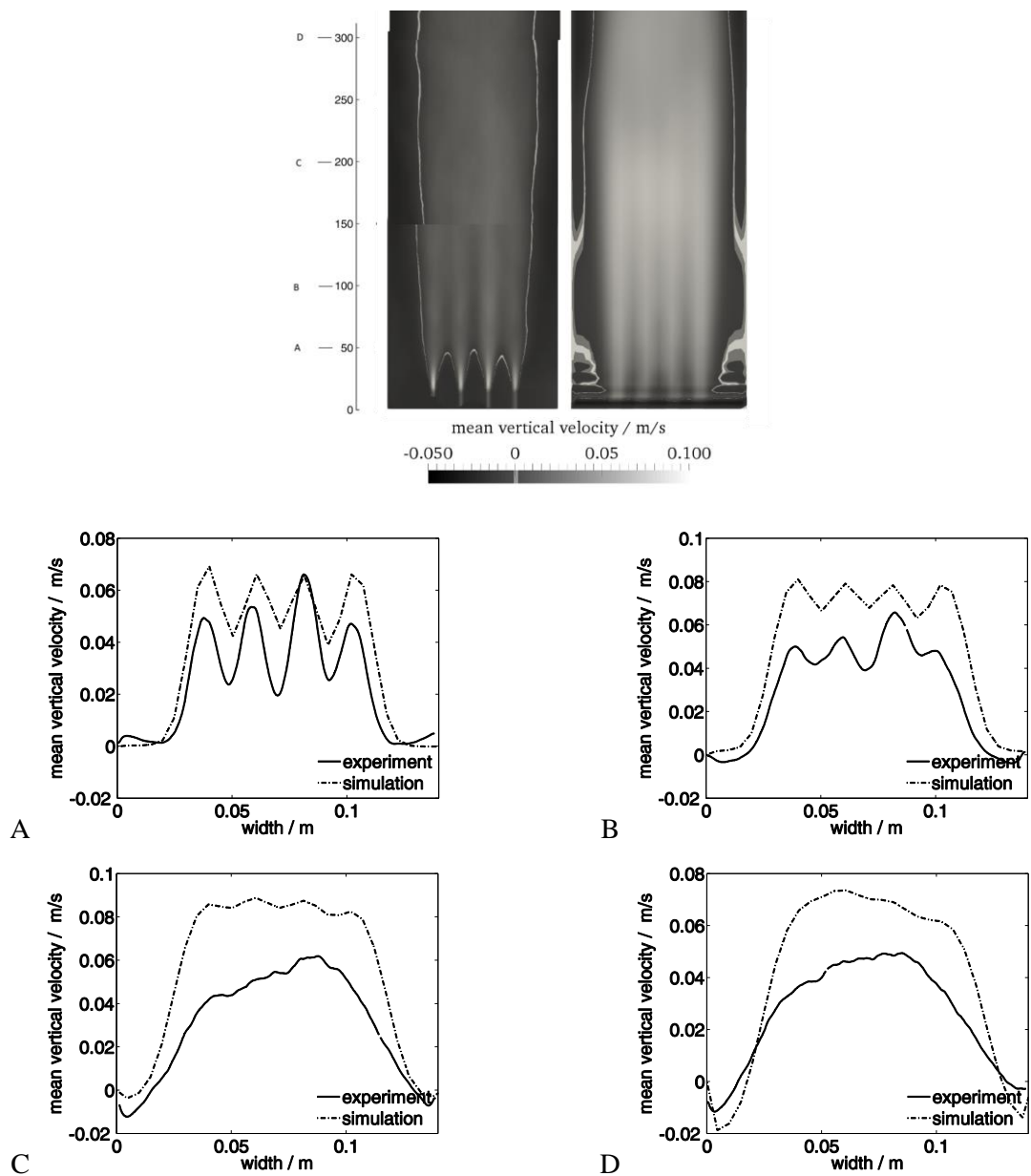

Figure 7: Mean vertical velocity, upper fig.: overview (left: experiment, right: simulation) A-D depicts the detailed line plot positions, A- D: Detailed line plots on different heights

Calculation of the simulated fluctuating velocity is made using the k-epsilon turbulence model. Values descripted here (s. Figure 8) are solution of the turbulent energy *k* with assumed isotropic turbulence velocities:

$$|\vec{u}'| = \sqrt{\frac{2}{3}k} \tag{35}$$

The upper figure shows the fluctuating velocity deviation of experiment (left) and simulation (right) on the middle plane of the column. Again, the scale on the left side shows the line plot positions A-D. In the lower sections, the fluctuation velocity is similar to the measured values from the experiments, in higher sections the simulation underestimates the level of turbulence, leading to low fluctuating velocities in comparison to experimental data. Here again, a standard parameter set was used for the BIT model, which was not optimized on this particular case as investigations on bubble induced turbulence was also not in the focus. However, as a consequence the impact of turbulence on bubble rotation is adjusted via the parameter study.

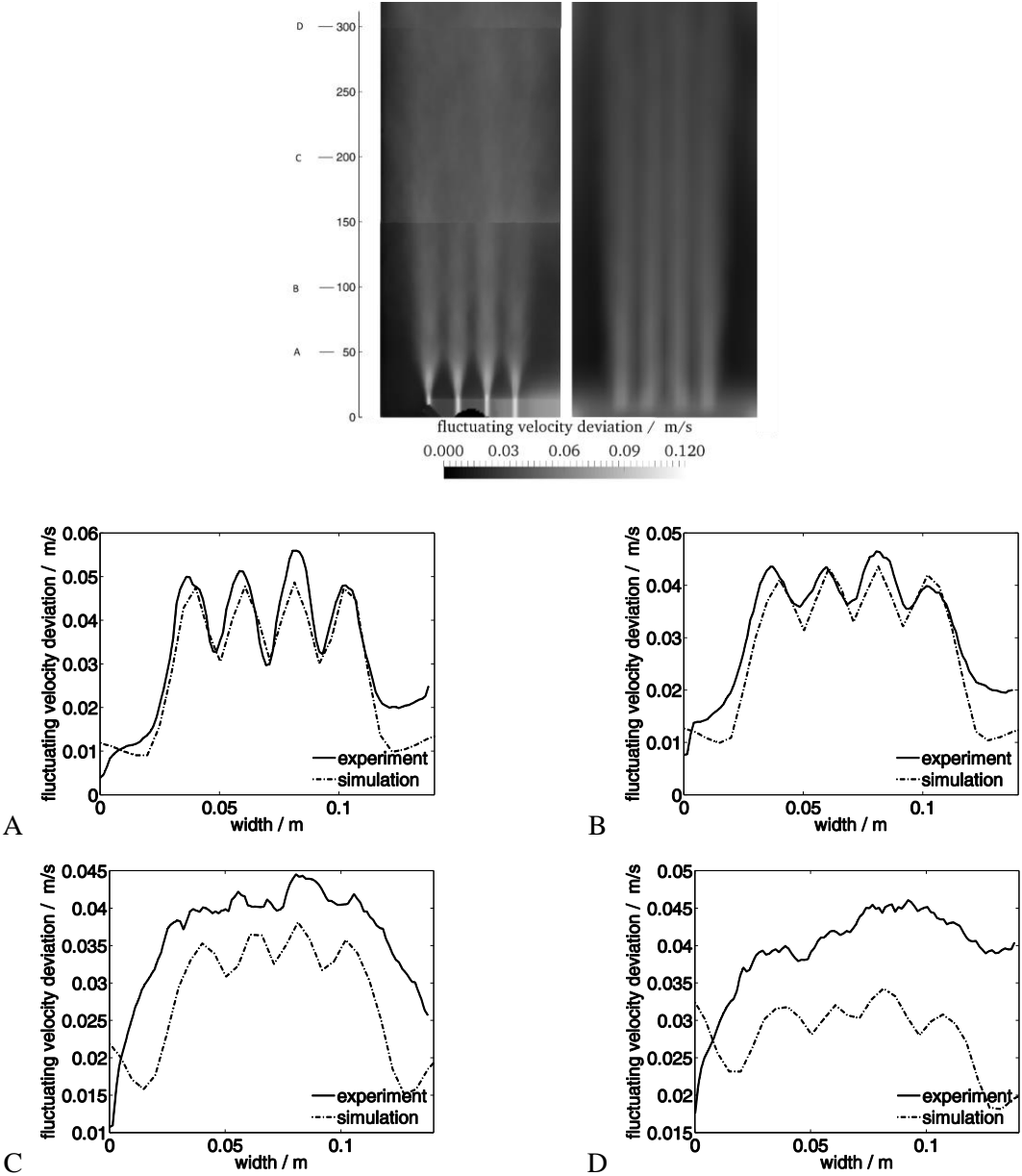

Figure 8: Standard deviation of velocity fluctuations, upper fig.: overview (left: experiment, right: simulation) A-D depicts the detailed line plot positions, A-D: Detailed line plots on different heights

*Bubble path characteristics*

According to literature (Shew & Pinton, 2006a), the bubble path oscillations of a single air bubble ($d$ = 3 mm) rising in tap water is approximately $f$ = 5 s$^{-1}$ and with its mean free rising velocity of $u_b$ = 35 cm/s the bubble path wavelength is $\lambda$ = 7 cm. According to eq. (32) this would recommend setting the parameter $\bar{\gamma}J_3 \approx 1000$.

Anyhow, analysis of the experimental data from the University of Magdeburg yielded a mean bubble path wavelength of $\lambda$ = 4.14 cm and an amplitude of $A$ = 1.87 mm. Mean bubble rise velocity is $u_b$ = 29 cm/s in this case which gives a frequency of $f$ = 7 s$^{-1}$ and a product of parameters of $\bar{\gamma}J_3 \approx 2000$. This was only a first estimation, since the random rotation is not be set to zero in the final simulation, which changes the frequency slightly.

Experimental data from 4 different measurements on the same properties were used to reconstruct and analyze bubble trajectories (s. Table 2). The trailing numbers in the Case name stand for the frame rate and the number of frames used in the analysis. A given frame rate of 0.1 kHz and 100,250 or 500 frames implies a measuring time of 1, 2.5 or 5 seconds accordingly. Deviation of wavelength and especially of amplitude is quite high, which is partly owed to the 2D analysis of the bubble paths. Since a flat zigzag path can only be seen correctly from one perspective, the calculated 2D amplitude will hardly correspond to the true 3D amplitude, it will generally underestimate the true value. This is why the standard deviation of the analyzed amplitude show rather high values in all experimental and simulation cases.

Table 2: *characteristic path lengths (all values in mm)*

| Case name | mean wavelength $\lambda$ | std. dev. wavelength | mean amplitude $A$ | std. dev. amplitude |
|---|---|---|---|---|
| Air V=10 l/h 0.1khz 1-100 | 40.6024 | 12.5520 | 1.8762 | 1.2405 |
| Air V=10 l/h 0.1khz 1-250 | 41.2272 | 13.1363 | 1.8810 | 1.2360 |
| Air V=10 l/h 0.1khz 1-500 | 41.7505 | 13.2859 | 1.8777 | 1.1981 |
| Air V=10 l/h 0.1khz 251-500 | 42.0848 | 13.4051 | 1.8642 | 1.1510 |
| Experiments mean value | 41.4162 | 13.0948 | 1.8747 | 1.2064 |
| Simulation 1 | 41.3764 | 14.6372 | 1.8069 | 1.1588 |
| Simulation 2 | 41.0091 | 14.6410 | 1.7593 | 1.1135 |
| Simulation 3 | 41.0273 | 14.6988 | 1.7472 | 1.1651 |

With the appropriate set of parameters, the simulated bubble path characteristics properly match the experimental values. Final simulations show that the assumed parameter $\bar{\gamma}J_3 \approx 1000$ to 2000 is a quite good starting value, the most promising results were done with parameters in the range of $\bar{\gamma}J_3 = 1250$ to 1350. It turned out that the values of the parameters for direct influence of liquid rotation ($\gamma J_1$) and shear ($\gamma J_2$) are very low in contrast to the oscillation ($\gamma J_3$), while the random factor ($R$) is of the same magnitude. Thus, the impact of rotational/shear flow around bubbles plays a minor role on the rotation while it is dictated mostly by turbulent eddies hitting the bubbles.

Table 3: *parameter sets for the three most promising simulations*

| Case name | $\bar{\gamma}J_3$ | $R$ | $\beta$ | $\bar{\gamma}J_1, \bar{\gamma}J_2$ |
|---|---|---|---|---|
| Simulation 1 | 1300 | 1950 | 0.15 | 32.50 |
| Simulation 2 | 1350 | 1620 | 0.15 | 33.75 |
| Simulation 3 | 1250 | 1875 | 0.15 | 31.25 |

A visual comparison of the three most promising bubble paths in Figure 9 reveal only few evident differences. The overall shape and distribution of path lines is almost identical. Most noticeable differences can be observed directly at the bubble inlets at the very bottom of the column. In the simulation bubbles are spread earlier than it is observed in the experiments. In the experiment bubbles describe a straight line directly after being injected to the column, while the simulation shows a rotation of the bubbles at the very beginning of the path, pushing the bubbles into a turn after the injection. In the first 100 mm a pattern can be seen in the flow paths of the simulation. The flow in the middle of the column pushes the bubbles away from the middle line, creating an area where almost no bubble cross. Some bubbles tend to take the same path multiple times, while in the experiment a slightly more chaotic distribution is present. In the simulation, a parameter for random rotation is used, but the analysis does not concern the overall distribution of bubbles. The spread/diffusion of bubbles in the upper area however (height > 150 mm) yields no (qualitatively) visible difference between experiment and simulation.

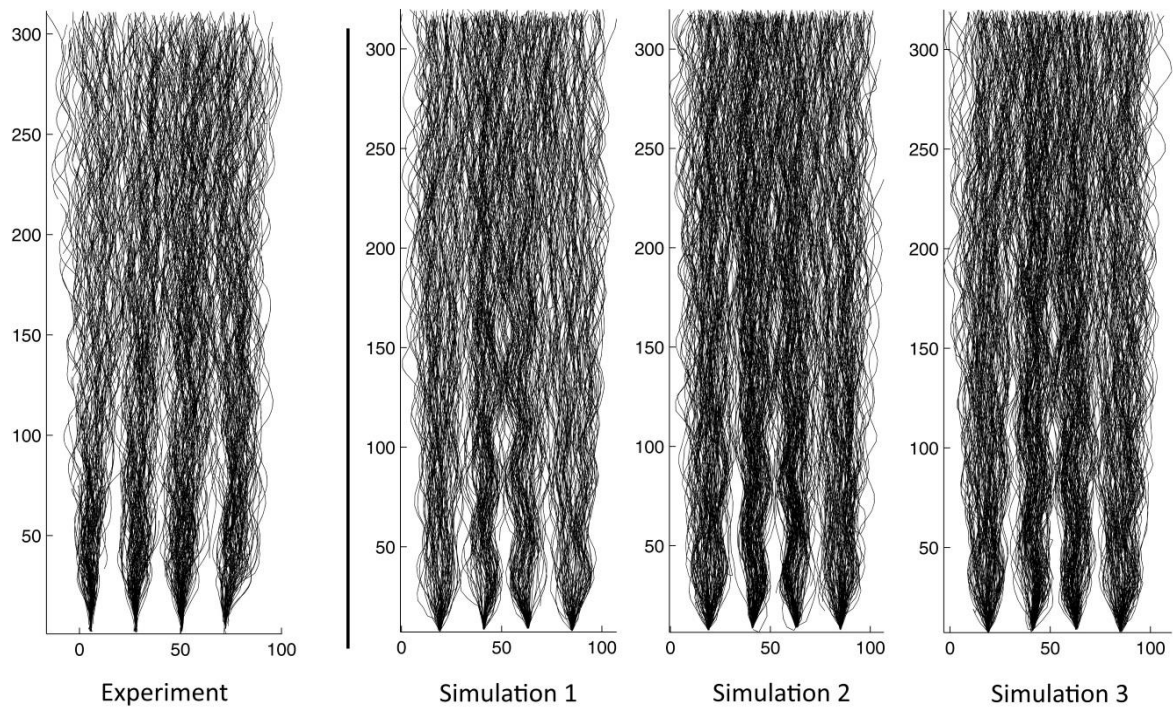

Figure 9: Reconstructed bubble trajectories, axes in mm, left: experimental results, right: simulation results

Another point of concern is the behavior of the simulated orientation in comparison with the velocity of the bubble. In order to describe this relations, different deduced angles are used; the movement angle lies between the vertical axis and movement direction, the orientation angle is between the short axis direction and vertical axis and the drift angle lies between orientation and movement vectors. With 2D analysis of the simulation data, the orientation angles show a normal distribution with mean value near to zero (Figure 10, left), just like the experimental results (Sommerfeld & Bröder, 2009). A 3D analysis however proves the oversimplification when using 2D analysis because the simulated mean 3D orientation is 18.3°. This gets clear when figuring a perfectly spiraling path seen from only one side, where there is a constant orientation angle in reality but an alternating angle is seen from a fixed observer. In a worst case scenario, a flat sinusoidal path oscillates parallel to the line of sight, thus making it impossible to see any oscillations at all.

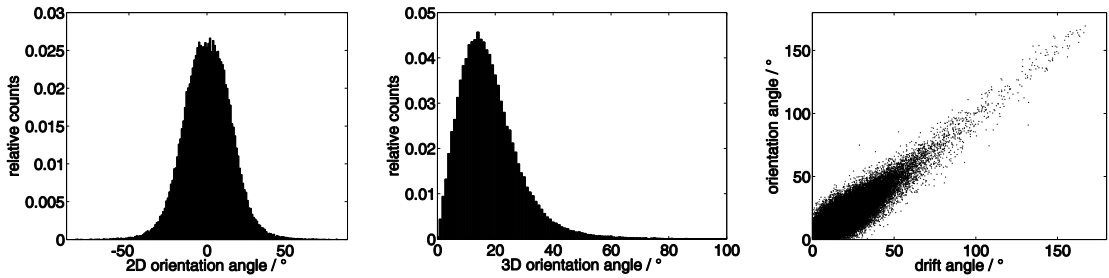

Figure 10: Characteristic angles between velocity and orientation vectors of simulated bubbles, left: orientation angle using 2D data, middle: orientation angle using 3D data, right: drift angle vs. orientation angle (3D data)

Bubble velocity and orientation vectors roughly point to the same direction, this has been shown in several references (Sommerfeld & Bröder, 2009; Ellingsen & Risso, 2001; Mougin & Magnaudet, 2006), which implies a small drift angle. Simulation results also show mostly small drift angles (Figure 10, right) with a mean value of 19.2°. Experimental data shows, that the largest drift angles occur at the most outer points of the oscillation, especially when a planar zigzag path is fulfilled (Mougin & Magnaudet, 2006). This means a high drift occurs when the orientation angle is near zero. Simulation data shows other characteristics; a high drift angle correlates with a high orientation angle (s. Figure 10, right). In DNS simulation, the drift and orientation angles have shown sinusoidal characteristics, while both were 90° out of phase. In the EL simulation these angles are in phase. Unfortunately, there is no experimental proof for this correlation with 3D measurements in a bubble swarm.

When taking a closer look at single trajectories of one of the bubbles rising in the swarm it will reveal characteristic parts of the bubble's movement. As shown in Ellingsen & Risso (2001), the bubble paths can be described as 'flattened helixes' which become less flattened while rising. Exactly this behavior can be found in the simulated bubble paths depicted in Figure 11. The bubble path shown here is not picked randomly, since characteristic movement is not achieved with every bubble. Note that this simulation considers bubbles in a large swarm rather than single rising bubbles (like in the experimental reference). This leads to a more chaotic liquid and bubble movement which can abruptly change bubble movement direction and orientation. Near the walls, a downward flow occurs in which bubbles performed differently, mostly not showing steady oscillations. As most bubbles rising in the middle part of the column, oscillation slowly starts within the lower 100 mm where it evolves to a sinusoidal and finally to a helical path (at approx. 400 mm height).

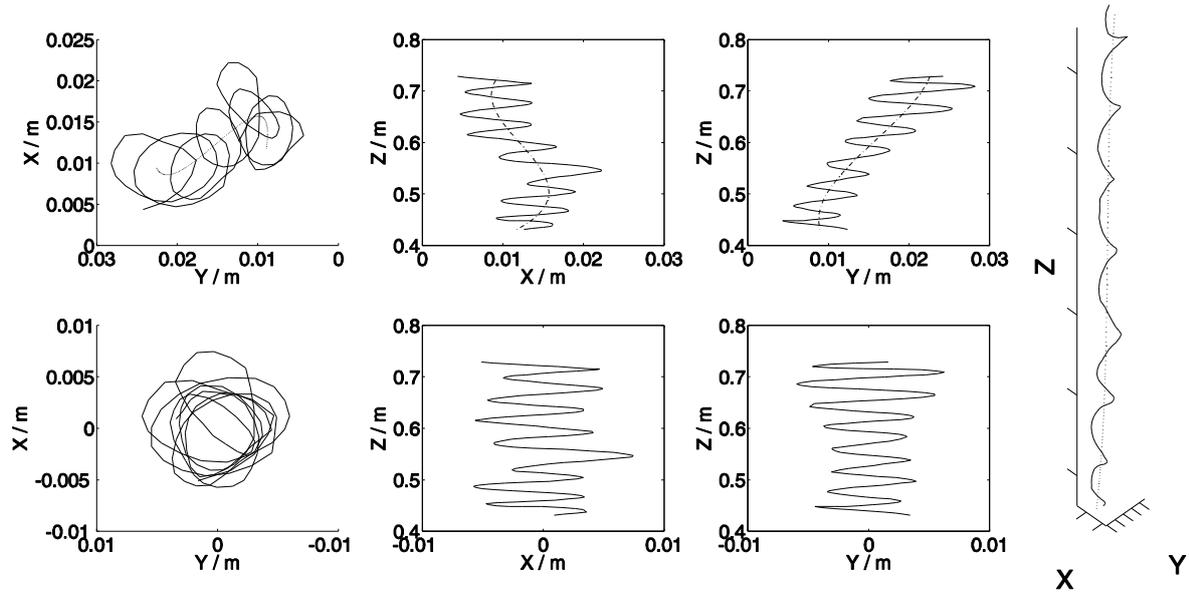

Figure 11: Single bubble trajectory; top row: raw data with fitted polynomials (dotted lines), bottom row: drift-corrected path data, right: isometric view of the raw path with fitted polynomials (dotted line)

We can observe influence of the flow in terms of a strong drift in the bubble paths (s. Figure 11, top row). In order to give a three dimensional description of the paths, a polynomial fit (n = 3) is used to smooth out the bubbles drift movement. In Figure 11 (top row) the original bubble paths are shown from three sides with the polynomial fit (dotted line). The bubble paths are normalized using the polynomial fit and the resulting path data (s. Figure 11, bottom row) describes a spiral, which is more comparable to experimental work (where the drift movement is almost zero). Assuming that our path describes a perfect flattened spiral, its coordinates perpendicular to the rising direction can be taken to calculate for a 'dynamic radius' of the path. A flattened spiral would look like an ellipsoid in the X-Y view (s. Figure 11, bottom left). The maximum of this radius will give the amplitude of the major oscillation mode, while the minimum will give the minor modes amplitude. When the minor amplitude is zero, a completely flat spiral, a planar zigzag path, is present. Also the wavelength in rising direction can be extracted from this data by measuring the distance for one spiral rotation. In Simulation case 1 the mean (3D) wavelength is $\lambda = 4.79$ cm and mean amplitude is $A = 4.04$ mm. This is a much larger amplitude, than it has been calculated for the 2D analysis, but comparison with Ellingsen & Risso (2001) shows a good match. They analyzed a single rising bubble with equivalent diameter $d = 2.48$ mm and could measure a major mode amplitude of $A = 4.3$ mm. Measured path frequency was $\omega = 39$ rad/s, which is $\omega = 35.4$ rad/s in our simulation case (mean rise velocity $u_b = 0.264$ m/s).

**Conclusions**

The presented EL simulation is capable of simulating unstable sinusoidal/spiraling bubble paths using macroscopic models. Bubble orientation, rotation and shape are calculated to achieve characteristic movement. Due to the assumption of bubbles describing rotational spheroids, the additional parameters that have to be calculated reduce to a shape factor, rotation and orientation vectors. A force induced by the bubbles rotation produces the lateral force leading to an oscillation movement.

A parameter study was used to fit the model constants to experimental data for mean bubble size of $d = 3$ mm. Characterization of the bubble paths was made using the amplitude and wavelength of the typical spiral movement. Evaluation of the 3D bubble path in a bubble swarm is difficult, since most references only supply 2D camera setups or single bubble trajectories in 3D analysis systems. It turned out, that most 2D measurements cannot reflect characteristic path parameters entirely, especially orientation angles are problematic. Nevertheless, the simulation results were mapped to a 2D point of view and compared to the experimental data. After parameter fitting, comparison to reference bubble path data was made using 2D and 3D analysis and could prove correct reproduction of unstable bubble paths. Comparison to DNS simulation and single rising bubble path data could also show good agreement. Amplitude and wavelength of the simulated bubble path are in unison with the measurements. Detailed comparison of DNS results of the drift angle reveal slight disagreement.

For further improvement of the model, a predictive parameter approach should be used to also cover different bubble sizes. Interaction of deformed bubbles are not considered in the EL model shown here, this could include collision, break-up, mass transfer and other shape dependent processes.

**Acknowledgment**